\documentclass[journal]{IEEEtran}
\usepackage{amsmath,amsfonts}
\usepackage{algorithmic}
\usepackage{algorithm}
\usepackage{array}
\usepackage{textcomp}
\usepackage{stfloats}
\usepackage{url}
\usepackage{verbatim}
\usepackage{graphicx}
\usepackage{cite}
\usepackage{multirow,tabularx}
\usepackage{booktabs,url}
\usepackage{subfigure,enumitem,setspace,color} 
\usepackage[table]{xcolor}
\usepackage{caption}
\usepackage{makecell}
\usepackage[font=footnotesize]{subfig}
\definecolor{gray}{rgb}{0.5,0.6,0.7}
\ifCLASSOPTIONcaptionsoff
  \usepackage[nomarkers]{endfloat}
 \let\MYoriglatexcaption\caption
\renewcommand{\caption}[2][\relax]{\MYoriglatexcaption[#2]{#2}}
\fi
\begin{document}

\title{UNQA: Unified No-Reference Quality Assessment for Audio, Image, Video, and Audio-Visual Content}

\author{Yuqin~Cao,~\IEEEmembership{Student Member,~IEEE,}
        Xiongkuo~Min,~\IEEEmembership{Member,~IEEE,}
        Yixuan~Gao,~\IEEEmembership{Student Member,~IEEE,}
        Wei~Sun,~\IEEEmembership{Member,~IEEE,}
        Weisi~Lin,~\IEEEmembership{Fellow,~IEEE,}
        and~Guangtao~Zhai,~\IEEEmembership{Senior Member,~IEEE}
\thanks{Y. Cao, X. Min, Y. Gao, W. Sun, and G. Zhai are with the Institute of Image Communication and Network Engineering, Shanghai Key Laboratory of Digital Media Processing and Transmissions, Shanghai Jiao Tong University, Shanghai
200240, China (e-mail: caoyuqin@sjtu.edu.cn; minxiongkuo@sjtu.edu.cn; gaoyixuan@sjtu.edu.cn; sunguwei@sjtu.edu.cn; zhaiguangtao@sjtu.edu.cn).

Weisi Lin is with the School of Computer Science and Engineering, Nanyang Technological University, Singapore 639798 (e-mail: wslin@ntu.edu.sg).
}}



\maketitle

\begin{abstract}
As multimedia data flourishes on the Internet, quality assessment (QA) of multimedia data becomes paramount for digital media applications. Since multimedia data includes multiple modalities including audio, image, video, and audio-visual (A/V) content, researchers have developed a range of QA methods to evaluate the quality of different modality data. While they exclusively focus on addressing the single modality QA issues, a unified QA model that can handle diverse media across multiple modalities is still missing, whereas the latter can better resemble human perception behaviour and also have a wider range of applications. In this paper, we propose the \underline{U}nified \underline{N}o-reference \underline{Q}uality \underline{A}ssessment model (\textbf{UNQA}) for audio, image, video, and A/V content, which tries to train a single QA model across different media modalities. To tackle the issue of inconsistent quality scales among different QA databases, we develop a multi-modality strategy to jointly train UNQA on multiple QA databases. Based on the input modality, UNQA selectively extracts the spatial features, motion features, and audio features, and calculates a final quality score via the four corresponding modality regression modules. Compared with existing QA methods, UNQA has two advantages: 1) the multi-modality training strategy makes the QA model learn more general and robust quality-aware feature representation as evidenced by the superior performance of UNQA compared to state-of-the-art QA methods. 2) UNQA reduces the number of models required to assess multimedia data across different modalities. and is friendly to deploy to practical applications. 
\end{abstract}

\begin{IEEEkeywords}
Quality assessment, multi-modality, joint training.
\end{IEEEkeywords}

\section{Introduction}
\begin{figure}[!tb]
\captionsetup[subfigure]{justification=centering}
\centering
  \includegraphics[width=0.9\linewidth]{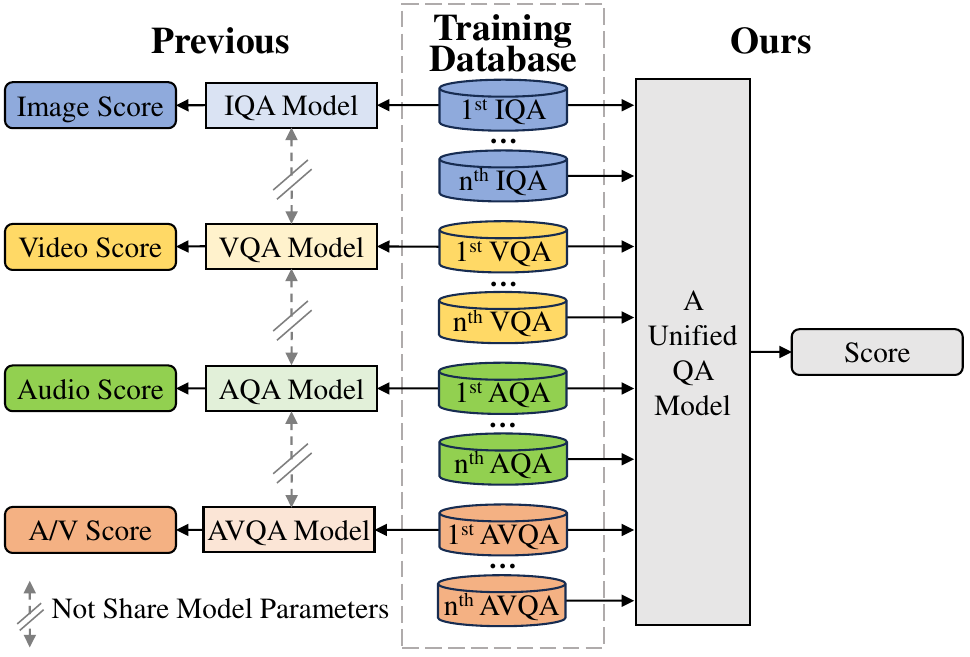}
  \caption{Most of previous QA models were trained on one QA database to measure the perceptual quality of a single modality. In this paper, we propose the first unified QA model that is jointly trained on multiple QA databases to predict qualities for inputs of different modalities. }
  \vspace{-0.3cm}
\label{fig:first}
\end{figure}
Recent years have witnessed the prosperity and development of multimedia data on the Internet, especially for user-generated content (UGC) on various social media and streaming media applications. However, various factors will introduce distortions to multimedia data, such as the non-professional equipment and shooting skills of amateur users, compression technology, and transmission systems. More and more service providers pay attention to quality assessment (QA) of multimedia data, which is crucial for multimedia processing and recommendation systems, such as filtering multimedia data of extremely low quality, recommending high-quality ones to users, guiding compression and transmission algorithms to achieve a trade-off of quality and bit rates, and so on.

Researchers have proposed various kinds of no-reference (NR) QA methods including audio quality assessment (AQA) \cite{lo2019mosnet,catellier2019wenets,fu2018quality,zezario2020stoi,li2013non,cao2023subjective}, image quality assessment (IQA) \cite{mittal2012making,6272356,min2017blind,min2017unified,sun2022blind,su2020blindly,gao2022image,min2021screen,min2020metric}, video quality assessment (VQA) \cite{Sun2021,li2021unified,sun2022deep,wu2022fast}, and audio-visual quality assessment (AVQA) \cite{cao2021deep,min2020study,Martinez2018}. In our previous work \cite{cao2021deep}, we have explored NR QA for UGC audio-visual (A/V) content. NR QA does not rely on reference information and has a wider range of applications, especially for UGC QA. However, similar to other existing NR QA methods, we only focus on single modality QA and train a database-specific model on the specific QA database as shown in the left half of Fig. \ref{fig:first}. We refer to these kinds of QA models as modality-specific QA models. Although they have achieved state-of-the-art (SOTA) performance, a unified QA model to predict the quality of audio, image, video, and A/V content simultaneously is still missing. Since AQA, IQA, VQA, and AVQA all strive to model the human perception system, they can be compatible and mutually beneficial in a unified QA model. Furthermore, a unified QA model can handle more complicated and volatile scenarios, such as when multimedia data lacks some modalities \cite{ying2022telepresence}.

\begin{figure*}[!tb]
\captionsetup[subfigure]{justification=centering}
\centering
   \includegraphics[width=\linewidth]{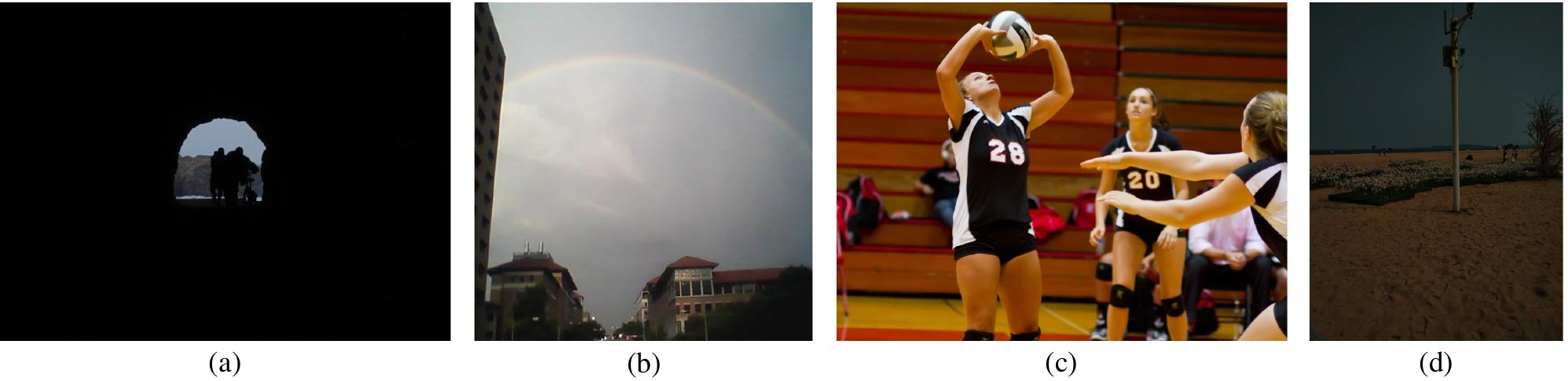}
   \caption{Images with approximately the same linearly rescaled MOS exhibit different perceptual quality. These images are sampled from (a) BID \cite{ciancio2010no}, (b) CLIVE \cite{ghadiyaram2015massive}, (c) KonIQ-10k \cite{hosu2020koniq} (d) SPAQ \cite{fang2020perceptual}. It is not hard to observe that the image (c) has the better quality than the other three.}
\label{fig:Image}
\end{figure*}

How to train a unified QA model across multiple QA databases simultaneously is a problem. The most direct way is to conduct joint training directly on multiple AQA databases, IQA databases, VQA databases, and AVQA databases. Due to different subjective testing methodologies, environments, and subjects, different QA databases have different perceptual scales. For example, the BID database \cite{ciancio2010no} was conducted the subjective experiment in a well-controlled laboratory environment with subjective scores in the range of $[0,5]$. The LIVE Challenge database \cite{ghadiyaram2015massive} was conducted in an unconstrained crowdsourcing platform with subjective scores in the range of $[0,100]$. This means that we need to conduct a separate subjective experiment to readjust the perceptual scale of images or videos in each database. To explain this more clearly, we linearly rescale the subjective scores of four IQA databases \cite{ciancio2010no,ghadiyaram2015massive,hosu2020koniq,fang2020perceptual} to $[0, 100]$ and show four images with approximately the same rescaled mean opinion scores (MOSs) in Fig. \ref{fig:Image}. We can observe that they have different perceptual qualities. 

To address this problem and develop a unified QA model, we make three contributions in this paper. Firstly, we propose \textbf{a \underline{U}nified \underline{N}o-reference \underline{Q}uality \underline{A}ssessment model} (UNQA) network that can flexibly predict quality scores for audio, image, video, and A/V content. Depending on the input modality, UNQA first selectively utilizes the spatial feature extraction, motion feature extraction, and audio feature extraction modules to derive the corresponding quality-aware features. Since visual-spatial information is of paramount importance for the human perception system, we utilize the multi-head self-attention (MHSA) in the spatial feature extraction module to capture salience information and guide the spatial feature extraction. Then, we fuse the quality-aware features via the corresponding modality-specific regression modules. UNQA can provide individual QA for each modality as well as overall quality prediction. These can be valuable for adjusting network traffic priorities for each modality and handling the ``missing-modality" problem \cite{ying2022telepresence}.

Secondly, we propose the first multi-modality training strategy for joint training on multiple AQA, IQA, VQA, and AVQA databases. While the absolute quality values of single modality content may not be directly comparable, we observe a consistent alignment in quality rankings among the audio/image/video/audio-visual pairs across different databases, so QA can be regarded as a learning ranking problem. Based on this, our multi-modality training strategy utilizes the relative ranking information from MOSs to bypass the problem that different databases have different perceptual scales. The strategy we proposed does not need to adjust the hyperparameter for each combination of joint training databases or increase the overall training cost, as the total number of training steps does not exceed the sum of all single-database training steps.

Thirdly, we utilize our multi-modality training strategy to jointly train UNQA on three AQA databases, four IQA databases, three VQA databases, and two AVQA databases, simultaneously. The experimental results show that joint training on various QA databases can make UNQA learn more general and robust feature representation and achieve better performance than SOTA QA models. In comparison to separate QA models for different modalities (AQA, IQA, VQA, and AVQA), UNQA combines their functions, leading to a more efficient use of storage space. It has practical advantages when edge devices, like smartphones and embedded systems, face memory constraints and cannot accommodate the weights of multiple QA models.

The rest of this paper is organized as follows. Section II introduces the related works. Then, we introduce our proposed unified QA model UNQA in Section III. Our multi-modality training strategy is described in Section IV. The experimental results are laid out in Section V. Section VI summarizes the paper.

\section{Most Related Work}
\subsection{Joint Training on Multiple QA Databases}
The challenge for joint training on multiple QA databases is that different databases have different perceptual scales. To bypass this problem, researchers consider pair-wise learning \cite{hu2019pairwise} for training on multiple databases. They first randomly sample pairs of images and feed image pairs into models. Then, they utilize the quality difference between image pairs as the ground truth to optimize the models. For pair-wise learning, different loss functions are proposed. For example, Yang~\textit{et al.}~\cite{yang2019cnn} utilized the margin ranking loss \cite{liu2017rankiqa} and the Euclidean loss. The ranking loss facilitates training with image pairs, aiming to determine which of the two images exhibits superior quality. Zhang~\textit{et al.}~\cite{zhang2021uncertainty} utilized the fidelity loss \cite{tsai2007frank} to train IQA models. However, pair-wise learning will increase the training cost. Li~\textit{et al.}~\cite{li2021unified} proposed a dataset-specific perceptual scale alignment for joint training on multiple VQA databases and did not increase extra training cost. Sun~\textit{et al.}~\cite{sun2023blind} proposed an iterative mixed database training strategy (IMDT) to train the model on multiple IQA databases simultaneously and train the regression modules on the corresponding target databases. While these studies only try to jointly train on single-modality QA databases, we propose the first multi-modality training strategy which jointly trains the unified QA model across QA databases of different modalities and does not increase the overall training cost. 

\subsection{Unified Multi-Modal QA Model}
The unified model aims to handle multi-modality inputs or multi-tasking whilst sharing parameters and computation. Some researchers \cite{girdhar2022omnivore,likhosherstov2021polyvit,girdhar2022omnimae} have proposed unified models that excel at classifying different visual modalities using exactly the same model parameters. In the field of QA, Ying~\textit{et al.}~\cite{ying2022telepresence} proposed a TVQA algorithm that extracted features from patches, frames, video clips, and audio clips from a video stream. It can provide IQA, VQA, and AVQA measurements for telepresence videos, respectively. They trained a database-specific model on the specific QA database and did not train on multiple QA databases, simultaneously. In this paper, we build UNQA which is jointly trained on multiple QA databases, including AQA, IQA, VQA, and AVQA databases, and predicts quality scores for different modalities using the same model parameters. UNQA leads to performance improvements on each individual QA database.
\begin{figure*}[!tb]
\captionsetup[subfigure]{justification=centering}
\centering
   \includegraphics[width=\linewidth]{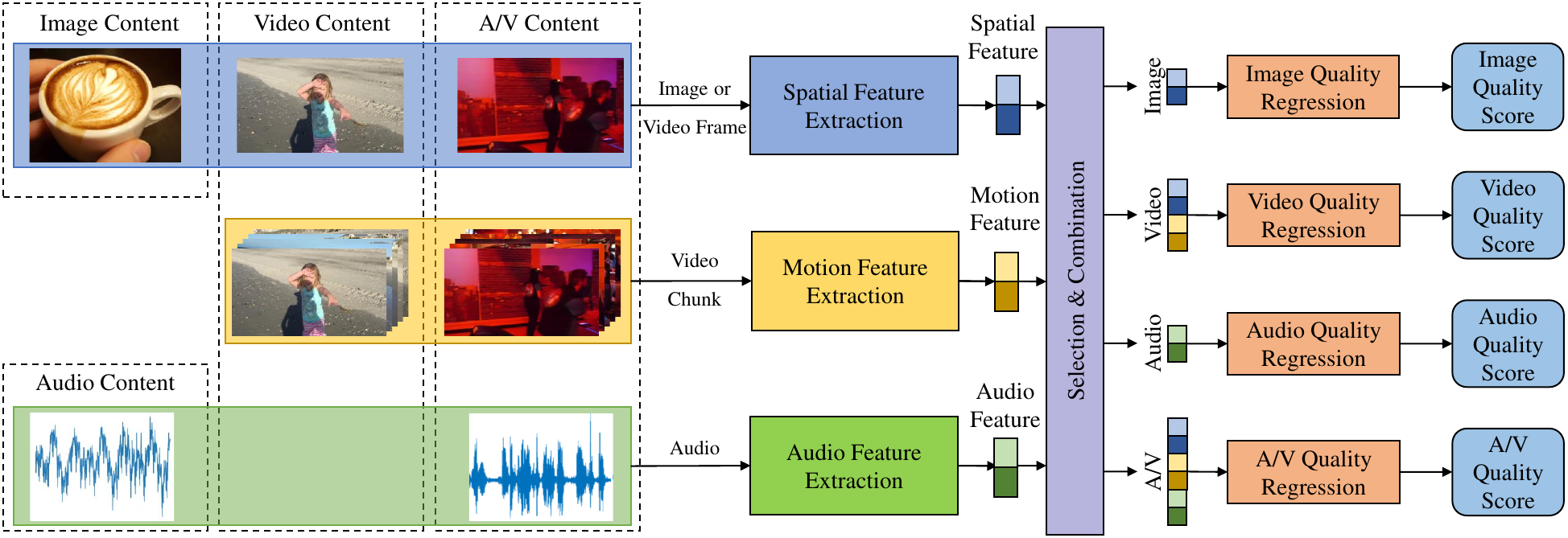}
   \caption{The overall architecture of the proposed model - Unified No-Reference Quality Assessment Model (UNQA). Our model is able to predict quality scores for different modalities: audio, image, video, and A/V. The model consists of three feature extraction modules and four modality-specific regression modules with minimal modality-specific changes.}
\label{fig:model}
\end{figure*}
\subsection{QA as Ranking}
Since most subjects give similar quality rankings when watching the same images or videos, QA can be seen as a learning-to-rank problem. We can derive the relative ranking information from MOSs to overcome the challenge posed by varying perceptual scales in different databases. Ma~\textit{et al.}~\cite{ma2017dipiq} utilized binary ranking information from FR IQA methods to train NR IQA models. While FR methods can only be applied to synthetic distortions, where the reference images are available, it is hard to apply to authentic distortions. Liu~\textit{et al.}~\cite{liu2017rankiqa} utilized images of the same content but with distortions at different levels to obtain discrete ranking information and pretrain the IQA model. However, they did not explore the idea of combining multi-modality databases via ranking information. As a result, their methods only achieve reasonable performance on a limited number of image synthetic distortions. We take a step further to train the unified QA model on multiple IQA, VQA, and AVQA databases based on relative ranking information.


\section{Unified QA Model}
\begin{figure*}[!tb]
\captionsetup[subfigure]{justification=centering}
\centering
   \includegraphics[width=0.88\linewidth]{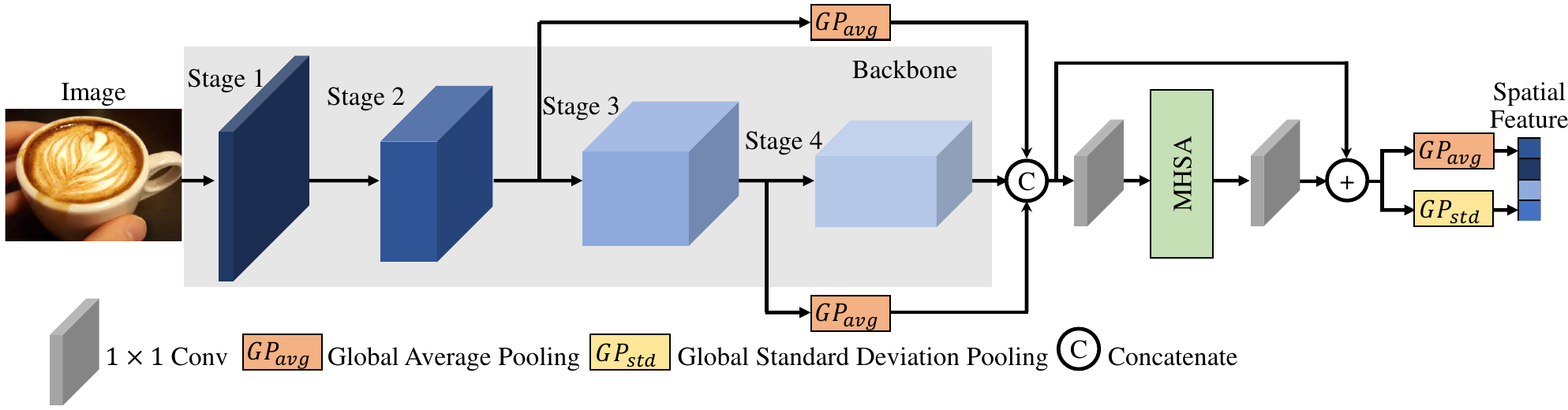}
   \caption{The architecture of the spatial feature extraction module. Assuming that the backbone model consists of $4$ stages, we extract feature maps from $3$ stages and feed them into MHSA to generate the spatial feature.}
\label{fig:SE}
\end{figure*}
In this section, we describe the architecture of our proposed model, UNQA, in detail. UNQA is designed to predict quality scores for audio, image, video, and A/V content. Since inputs from different modalities possess various dimensions, we utilize different feature extraction modules including spatial, motion, audio feature extraction modules to extract the corresponding dimensional features and regress them into a final quality score by four modality-specific quality regressors. The overall architecture of UNQA is illustrated in Fig. \ref{fig:model}.
\subsection{Spatial Feature Extraction Module}
In the spatial feature extraction module, we extract spatial features from images and video frames. Spatial features capture the spatial distortions introduced in images and video frames, such as noise, blur, block effect, overexposure, low light, etc. Previous studies have indicated that the convolutional neural network (CNN) models have excellent feature extraction ability \cite{sun2019mc360iqa,yang2019cnn,sun2023influence}. The features extracted by the shallow layers of models contain low-level information, such as edges, corners, textures, etc. The features extracted by the deep layers of the CNN models contain rich semantic features. We use a popular CNN model (i.e. ConvNext V2 \cite{woo2023convnext}) pre-trained on the ImageNet database \cite{R8} as the backbone to extract quality-aware features from both the shallow layers and deep layers. Then we combine them to generate the spatial features, so as to leverage their complementary advantages. The framework of the spatial feature extraction module is shown in Fig. \ref{fig:SE}.

We utilize a 5D tensor $X\in \mathbb{R}^{B\times T\times C \times H \times W}$ to represent the input of the spatial feature extraction module, where $B$ is the batch size, $T$ is the temporal dimension, $C$ is the color channel, and $H,W$ are the spatial dimensions. Images can be regarded as single-frame videos with $T=1$. We first convert input into a 4D tensor $X\in \mathbb{R}^{BT\times C \times H \times W}$, and then feed it into the backbone model. Assuming that the backbone model consists of $N$ stages, we extract the feature map $F_i$ from the $i$-th stage, where $F_i\in \mathbb{R}^{C_i\times H_i \times W_i}, i\in[1,2,...,N]$, and $C_i$, $H_i$, and $W_i$ are the channel, height, and width of the feature map $F_i$. Then, we apply global average operations on each feature map so that the height and width of each feature map are the same. After that, the feature map $\tilde{F}_i$ can be denoted as:
\begin{equation}
    \tilde{F}_i = \mathrm{GP_{avg}}(F_i), i\in[1,2,..,N-1],
\end{equation}
where $\tilde{F}_i\in \mathbb{R}^{C_i\times H_N \times W_N}$. Since MHSA can capture local salience information spatially, and generate multiple attention maps from different aspects, BoTNet \cite{srinivas2021bottleneck} combines spatial convolution and MHSA by replacing the convolutional layers with MHSA in the last three bottleneck blocks of ResNet. Inspired by them, we concatenate $\tilde{F}_i$ and feed them into MHSA to generate the feature map $\tilde{F}_{s}$:
\begin{equation}
    \tilde{F} =  \mathrm{cat}(\{\tilde{F}_i\}^N_{i=1}),
\end{equation}
\begin{equation}
    \tilde{F}_{s} = \tilde{F} + \mathrm{Conv}( \mathrm{MHSA}(\mathrm{Conv}(\tilde{F}))),
\end{equation}
where $\tilde{F}_{s}\in \mathbb{R}^{\sum\limits_{i=1}^{N}C_i\times H_N \times W_N}$ and $\mathrm{cat}(\cdot)$ is the concatenation operation. Then, we apply global average and standard deviation pooling operations on the feature map $\tilde{F}_{s}$:
\begin{equation}
    \mu =  \mathrm{GP_{avg}}(\tilde{F}_{s}),
\end{equation}
\begin{equation}
    \sigma =  \mathrm{GP_{std}}(\tilde{F}_{s}),
\end{equation}
where $\mu$ and $\sigma$ are the $1120$-dimensional global mean and standard deviation of feature map $\tilde{F}_{s}$ respectively. Finally, we concatenate the $\mu$ and $\sigma$ to represent the spatial feature $F_s$:
\begin{equation}
    F_s = cat(\{\mu, \sigma\}).
\end{equation}
\subsection{Motion Feature Extraction Module}
Besides spatial distortions, motion distortions also degrade the perceptual quality of video and A/V content. The unstable shooting equipment and the fast movement of the shooting object will lead to motion distortions. The spatial features extracted from the intra-frame make it difficult to capture motion distortions. Therefore, we utilize the pre-trained action recognition model as the motion feature extractor to obtain the motion feature of each video chunk:
\begin{equation}
    F_m = \mathrm{motion}(X_c),
\end{equation}
where $F_m$ represents the motion feature and $X_c$ represents the video chunk.
\begin{figure}[!tb]
\captionsetup[subfigure]{justification=centering}
\centering
  \includegraphics[width=0.7\linewidth]{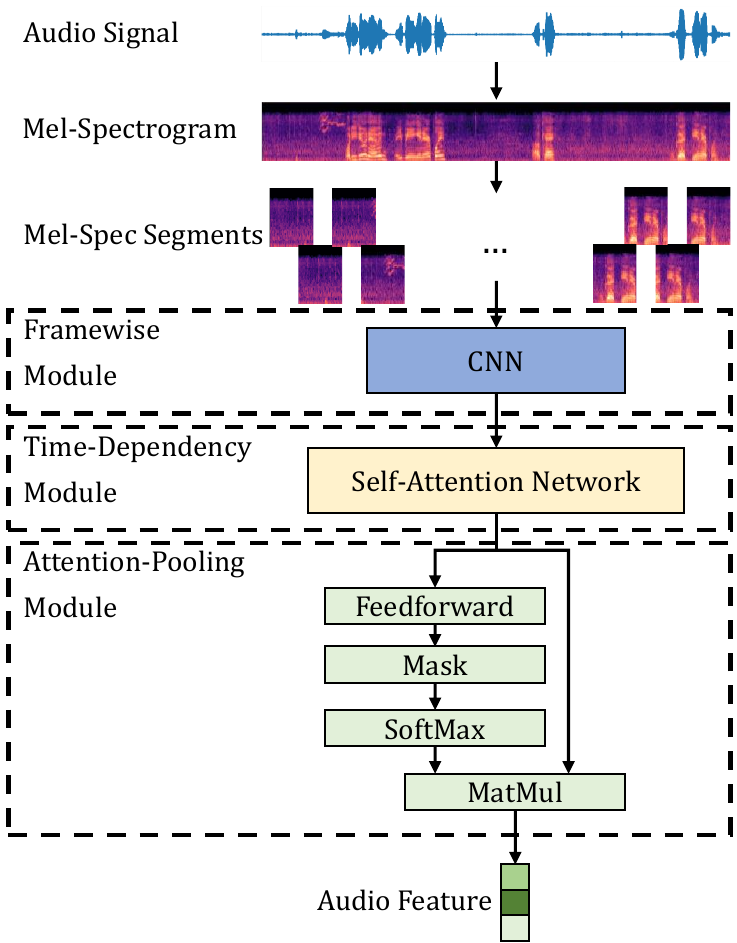}
  \caption{The architecture of the audio feature extraction module. It consists of a framewise module, a time-dependency module, and an attention-pooling module.}
\label{fig:AE}
\end{figure}
\subsection{Audio Feature Extraction Module}
In the audio feature extraction module, we extract the audio feature $F_a$ from the audio content and the audio of A/V content. The framework of the audio feature extraction module is shown in Fig. \ref{fig:AE}. We first convert one-dimensional audio signals into two-dimensional mel-spectrograms and then divide them into overlapping segments. Mittag~\textit{et al.}~\cite{mittag2021nisqa} proposed a deep CNN-self-attention model for multidimensional speech quality prediction which is focused on distortions that occur in communication networks. Inspired by them, we first utilize a CNN as the framewise module to extract features from mel-spectrograms, then model the time dependencies through a self-attention network, and finally aggregate features over time in the pooling module. The aggregated features are denoted as the audio feature of our model.
\subsection{Regression Module}
For inputs of different modalities, we need to utilize different feature extraction combinations to extract the corresponding features:
\begin{equation}
\begin{split}
    &F_{audio} = F_a,\\
    &F_{image} = F_s,\\
    &F_{video} = \mathrm{cat}(\{F_s, F_m\}),\\
    &F_{A/V} = \mathrm{cat}(\{F_s, F_m, F_a\}),
\end{split}
\end{equation}
where $F_{audio}$, $F_{image}$, $F_{video}$, and $F_{A/V}$ denote the feature of audio, image, video, and A/V, respectively. We fuse these features into the final quality score via four quality regression modules, including the audio quality regression module, image quality regression module, video quality regression module, and A/V quality regression module, which are all composed of two fully connected layers. 

\section{Multi-Modality Training Strategy}

\begin{figure*}[ht]
\centering
\includegraphics[width=\linewidth]{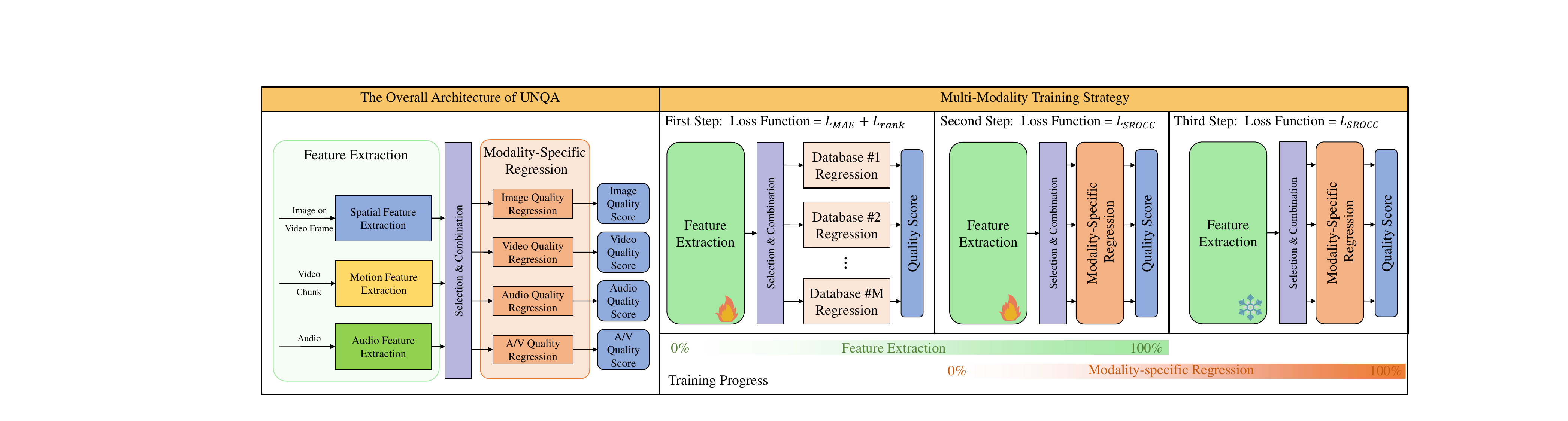}
\caption{The procedure of our multi-modality training strategy for UNQA which consists of three steps. (a) First Step: pretrain the three feature extraction modules. (b) Second Step: merge database-specific regression modules into three modality-specific regression modules. (c) Third Step: finetune the three modality-specific regression modules.}
\label{fig:Procedure}
\end{figure*}
We jointly train UNQA across multiple AQA, IQA, VQA, and AVQA databases. As shown in Fig. \ref{fig:Procedure}, our multi-modality training strategy can be divided into three steps. The details of each step are described as follows.

\subsection{First Step}
\begin{figure}[!tb]
\captionsetup[subfigure]{justification=centering}
\centering
  \includegraphics[width=\linewidth]{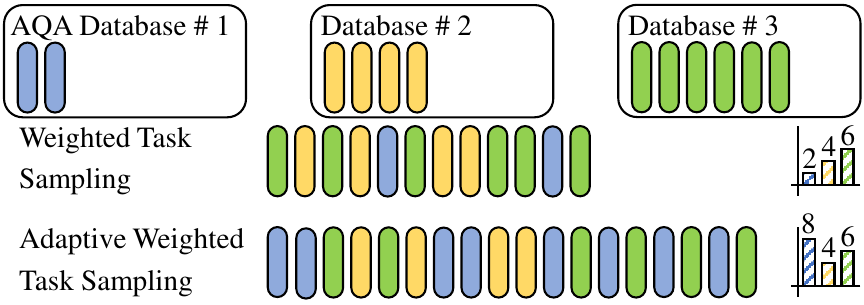}
  \caption{Overview of the weighted task sampling method.}
\label{fig:sampling}
\end{figure}
The first step aims to pretrain and align the spatial feature extraction module, motion feature extraction module, and audio feature extraction module. Assuming that we jointly train the model on $M$ QA databases, we replace four modality-specific regression modules with $M$ database-specific regression modules corresponding to each database and keep other modules unchanged. After the three feature extraction modules capture the general quality-aware features across different databases, the database-specific regression module, trained on the individual database, can adjust them to the corresponding database-specific perceptual scale. Therefore, UNQA can directly take the absolute quality scores as the ground truth in the first step. For each optimization step, we sample a minibatch from one database, calculate a gradient, and then perform a parameter update. An important consideration is the order in which we sample databases. Likhosherstov~\textit{et al.}~\cite{likhosherstov2021polyvit} compared five sampling schedules and found that the weighted task sampling method performs the best. The weighted task sampling method samples each database with a weight proportional to the number of training steps for each database. Since learning rates between visual and audio modalities are different, direct joint training can lead to overfitting or underfitting issues. To unlock the potential of joint training, we propose an adaptive weighted task sampling method to increase the sampling frequency of audio modality during training. As shown in Fig. \ref{fig:sampling}, assuming that the training steps of the AQA database, No.2 database, and No.3 database are $2$, $4$, and $6$, respectively. In the adaptive weighted task sampling method, we repeat the training steps for the AQA database, thereby increasing the training steps of the AQA database from $2$ to $8$. It makes the learning of audio modalities more harmonious with visual modalities. Inspired by \cite{sun2022deep}, we combine the mean absolute error (MAE) loss $L_{MAE}$ and rank loss $L_{rank}$ as the loss function to optimize the proposed model. The loss function can be defined as:
\begin{equation}
    L = L_{MAE} + L_{rank},
\end{equation}
where $L_{MAE}$ makes the predicted scores close to the ground truth and $L_{rank}$ makes the ranking of predicted scores similar to the ranking of ground truth. The $L_{MAE}$ and $L_{rank}$ are defined as:
\begin{equation}
    L_{MAE} = \frac{1}{B}\sum\limits^B_{i=1}|o_i-s_i|,
\end{equation}
\begin{equation}
    L_{rank} = \frac{1}{B^2}\sum\limits^B_{i=1}\sum\limits^B_{j=1}L_{rank}^{ij},
\end{equation}
\begin{equation}
    L_{rank}^{ij} = max(0,|s_i-s_j|-e(s_i,s_j) \cdot (o_i-o_j)),
\end{equation}
\begin{equation}
    e(s_i,s_j) =
    \left\{
    \begin{aligned}
    1, s_i \ge s_j, \\
    -1, s_i < s_j, \\
    \end{aligned}
    \right.
\end{equation}
where the $s_i$ and $o_i$ are the ground truth quality score and predicted score of the $i$-th sample in a mini-batch respectively, and $B$ denotes the batch size.

\subsection{Second Step}
The second step is to merge $M$ database-specific regression modules into four modality-specific regression modules. We utilize three feature extraction modules pre-trained in the first step to extract features and calculate final quality scores through the regression module of the corresponding modality. Similar to the first step, we adopt the adaptive weighted task sampling method to sample the minibatch from each database. Since databases with the same modality but different perceptual scales will utilize the same modality-specific regression module to predict final scores, we utilize the Spearman’s Rank-order Correlation Coefficient (SRCC) loss \cite{li2022blindly} to convert the QA problem into a learning-to-rank problem and bypass the perceptual scale problem. The SRCC loss function can be defined as:

\begin{align}
    &L_{SRCC} = 1 - SRCC,\\
    SRCC& = \frac{\sum_i(o_i^r-\overline{o^r})(s^r_i-\overline{s^r})}{\sqrt{\sum_i(o^r_i-\overline{o^r})^2\sum_i(s^r_i-\overline{s^r})^2}},
\end{align}

where $\{o^r_i\}^B_{i=1}$ and $\{s^r_i\}^B_{i=1}$ denote the ranks of the predicted scores $\{o_i\}^B_{i=1}$ and the ground truth quality scores $\{s_i\}^B_{i=1}$. 

\subsection{Third Step}
The third step aims to finetune the regression module. Four feature extraction modules are jointly trained on multiple AQA, IQA, VQA, and AVQA databases, while the modality-specific regression modules are only trained with the inputs of the corresponding modality. It may cause the parameter update speed of the feature extraction modules and the regression module to be out of sync. For example, after the spatial feature extraction module is optimized by the image inputs, the video quality regression module may not adapt to the updated parameters of the spatial feature extraction module. Therefore, we fix the parameters of three feature extraction modules and only train the parameters of the regression module. Similar to the second step, we adopt the SRCC loss function to train the model.

\section{Experiments}
We first introduce our experimental settings, including test databases, compared methods, evaluation criteria, and implementation details. Then, we compare our proposed model UNQA with SOTA NR QA methods. After that, we conduct ablation studies to validate the contributions of different components. We also analyze UNQA performance in the cross-database evaluation and provide interpretability for UNQA by Grad-CAM \cite{selvaraju2017grad}. Finally, we test the computational efficiency of UNQA.

\begin{table}[!t]
    \centering
    \caption{The basic information of three AQA and four IQA databases. VOIP: Voice-over-Internet Protocol}
    \small
    \resizebox{0.45\textwidth}{!}{
    \begin{tabular}{c| ccccc}
    \toprule
    Attribute & Number & Distortion Type & MOS Range\\
    \midrule
    ITU-T \cite{recommendation1998itu} & 1,328 & Noise, codec, packet loss & [1,5]\\
    NOIZEUS \cite{hu2006subjective} & 1,792 & Noise & [1,5]\\
    TCDVoIP \cite{harte2015tcd}& 384 & VoIP degradations & [1,5]\\
    \midrule
    BID \cite{ciancio2010no}& 586 & Authentic & [0,5]\\
    CLIVE \cite{ghadiyaram2015massive}& 1,162 & Authentic & [0,100]\\
    KonIQ-10k \cite{hosu2020koniq}& 10,073 & Authentic & [1,5]\\
    SPAQ \cite{fang2020perceptual}& 11,125 & Authentic& [0,100]\\
    \bottomrule
    \end{tabular}
    }
    \label{tab:AQA Database}
\end{table}
\begin{table}[!t]
    \centering
    \caption{The basic information of three VQA databases and two AVQA databases. FR: Framerate.}
    \small
    \resizebox{0.5\textwidth}{!}{
    \begin{tabular}{c| ccccc}
    \toprule
    Attribute & Number & Resolution & Length & FR & Distortion\\
    \midrule
    YouTube-UGC \cite{wang2019youtube}& 1,184 & 360p-4k & 20s & 15-60 & Authentic\\
    KoNViD-1k \cite{hosu2017konstanz}& 1200 & 540p & 8s & 24-30 & Authentic\\
    LIVE-VQC \cite{sinno2018large}& 585 & 480p-1080p & 10s & 20-30 & Authentic\\ 
    \midrule
    SJTU-UAV \cite{cao2023UGC}& 520 & 540p-1080p & 8s & 15-30 & Authentic\\
    LIVE-SJTU \cite{min2020study}& 336 & 1080p & 8s & 24-30 & Synthetic\\
    \bottomrule
    \end{tabular}
    }
    \label{tab:VQA Database}
\end{table}
\begin{table*}[!htb]
    \centering
    \caption{Performance comparison between UNQA and SOTA QA methods on IQA, VQA, AQA, and AVQA databases. UNQA is jointly trained on all QA databases to generate a comprehensive model. StairIQA is jointly trained on all IQA databases to obtain separate model parameters for each database, while UNQA-Single and the other QA methods are trained separately on each database to obtain database-specific models. The best and second-best performances for each metric are marked in boldface and underlined, respectively. The $^*$ means that the results are cited from the original paper.}
    \label{tab:all}
    \begin{spacing}{1.1}
    \resizebox{\textwidth}{!}{
    \begin{tabular}{c||cc cc cc cc||c||cc cc}
    \toprule
    \multirow{2}{*}{IQA Method} & \multicolumn{2}{c}{BID} & \multicolumn{2}{c}{CLIVE} & \multicolumn{2}{c}{KonIQ-10k} & \multicolumn{2}{c||}{SPAQ} & \multirow{2}{*}{AVQA Method} & \multicolumn{2}{c}{SJTU-UAV} & \multicolumn{2}{c}{LIVE-SJTU}\\
    ~  & SRCC & PLCC & SRCC & PLCC & SRCC & PLCC & SRCC & PLCC & ~ & SRCC & PLCC & SRCC & PLCC\\
    \midrule
    NIQE & 0.4902 & 0.4816 & 0.4508 & 0.5122 & 0.5218 & 0.5217 & 0.4950 & 0.4889 & Linear & 0.4051 & 0.3679 & 0.6261 & 0.5872\\
    BRISQUE & 0.5637 & 0.5851 & 0.5924 & 0.6138 & 0.7119 & 0.7177 & 0.7239 & 0.7270 & Power & 0.3983 & 0.3620 & 0.6269 & 0.5921\\
    CNNIQA & 0.5569 & 0.4490 & 0.5371 & 0.4318 & 0.6104 & 0.4625 & 0.7456 & 0.6747 & Minkowski & 0.3906 & 0.3596 & 0.6339 & 0.5949\\
    DBCNN & 0.8180 & 0.8541 & 0.8316 & 0.8566 & 0.8772 & 0.8952 & 0.9127 & 0.9179 & NR-DNFAVQ & 0.6413 & 0.6401 & 0.8824 & 0.8900\\
    HyerIQA & 0.8221 & 0.8558 & 0.8452 & 0.8686 & 0.8792 & 0.8971 & 0.9068 & 0.9105 & DNN-RNT & 0.6918 & 0.6781 & \underline{0.9363} & \underline{0.9267}\\
    StairIQA & \underline{0.8673} & \underline{0.9066} & \underline{0.8696} & \underline{0.8952} & 0.8703 & 0.8971 & 0.9072 & 0.9132 & DNN-SND & 0.6717 & 0.6952 & 0.9063 & 0.8897\\
    \midrule
    UNQA-Single & 0.8339 & 0.8586 & 0.8591 & 0.8850 & \textbf{0.9102} & \textbf{0.9269} & \underline{0.9141} & \underline{0.9186} & UNQA-Single & \underline{0.7824} & \underline{0.7935} & 0.9184 & 0.8983\\
    UNQA & \textbf{0.8812} & \textbf{0.9140} & \textbf{0.8744} & \textbf{0.9027} & \underline{0.8930} & \underline{0.9153} & \textbf{0.9127} & \textbf{0.9174} & UNQA & \textbf{0.8301} & \textbf{0.8379} & \textbf{0.9541} & \textbf{0.9309}\\
    \bottomrule
    \end{tabular}
    }
    \end{spacing}
    \begin{spacing}{1.1}
    \resizebox{\textwidth}{!}{
    \begin{tabular}{c||cc cc cc ||c||cc cc cc}
    \toprule
    \multirow{2}{*}{VQA Method} & \multicolumn{2}{c}{YouTube-UGC} & \multicolumn{2}{c}{KoNViD-1k} & \multicolumn{2}{c||}{LIVE-VQC} & \multirow{2}{*}{AQA Method} & \multicolumn{2}{c}{ITU} & \multicolumn{2}{c}{NOIZEUS} & \multicolumn{2}{c}{TCDVoIP}\\
    ~ & SRCC & PLCC & SRCC & PLCC & SRCC & PLCC & ~ & SRCC & PLCC & SRCC & PLCC & SRCC & PLCC\\
    \midrule
    V-BLIINDS & 0.5590$^*$ & 0.5551$^*$ & 0.7101$^*$ & 0.7037$^*$ & 0.6939$^*$ & 0.5078$^*$ &  MOSNet & 0.6296 & 0.6127 & 0.6754 & 0.6634 & 0.6019 & 0.5926\\ 
    VIDEVAL & 0.7719 & 0.7691 & 0.7930 & 0.7904 & 0.7367 & 0.7347 & Wenets & 0.7369 & 0.7321 & 0.8905 & 0.8879 & 0.5919 & 0.6649\\
    RAPIQUE & 0.7640 & 0.7727 & 0.8066 & 0.8203 & 0.7513 & 0.7685 & Quality-Net & 0.7786 & 0.7764 & 0.8779 & 0.8744 & 0.6998 & 0.7546\\
    VSFA & 0.7644 & 0.7612 & 0.7819 & 0.7895 & 0.6994 & 0.7366 & STOI-Net & 0.7945 & 0.7827 & 0.8782 & 0.8775 & 0.6940 & 0.6250\\
    SimpleVQA & 0.7962 & 0.7954 & 0.8306 & 0.8340 & 0.7486 & 0.7849 & NRMusic & 0.8205 & 0.8232 & 0.9055 & 0.9032 & 0.7113 & 0.7414\\
    FastVQA & 0.7891 & 0.7805 & \underline{0.8473} & \underline{0.8435} & 0.7458 & 0.7640 & ARN-50 & 0.8317 & 0.8440 & 0.8707 & 0.8683 & \underline{0.8327} & \underline{0.8453}\\
    \midrule
    UNQA-Single & \underline{0.8113} & \underline{0.8025} & 0.8263 & 0.8311 & \underline{0.8032} & \underline{0.8175} & UNQA-Single & \underline{0.8529} & \underline{0.8592} & \underline{0.9221} & \underline{0.9231} & 0.7955 & 0.8404\\
    UNQA & \textbf{0.8411} & \textbf{0.8457} & \textbf{0.8666} & \textbf{0.8669} & \textbf{0.8235} & \textbf{0.8351} & UNQA & \textbf{0.8565} & \textbf{0.8662} & \textbf{0.9225} & \textbf{0.9269} & \textbf{0.8416} & \textbf{0.8719}\\
    \bottomrule
    \end{tabular}
    }
    \end{spacing}
\end{table*}
\begin{table*}[!htb]
    \centering
    \caption{The total number of parameters of UNQA and deep learning-based QA models. Except for UNQA, other models are trained separately on each database to obtain multiple models. We report the total number of parameters across all models in the ``Total params'' row.}
    \label{tab:params}
    \resizebox{\textwidth}{!}{
    \begin{tabular}{c|cccc|ccc|ccc|c}
    \toprule
    \multirow{2}{*}{Method} & \multicolumn{4}{c|}{IQA} & \multicolumn{3}{c|}{VQA} & \multicolumn{3}{c|}{AVQA} & Mixed\\
    ~ & CNNIQA & DBCNN & HyerIQA & StairIQA & VSFA & SimpleVQA & FastVQA & NR-DNFAVQ & DNN-RNT & DNN-SND & UNQA\\
    \midrule
    {\#Database} & 4 & 4 & 4 & 4 & 3 & 3 & 3 & 2 & 2 & 2 &9\\
    {\#Models} & 4 & 4 & 4 & 4 & 3 & 3 & 3 & 2 & 2 & 2 & 1 \\
    Total params & 2.764M & 58.408M & 104.428M & 119.304M & 68.85M & 166.986M & 107.300M & 44.838M & 130.884M & 119.996M & 52.382M\\
    \bottomrule
    \end{tabular}
    }
\end{table*}
\begin{figure*}[!t]
 \centering
    \includegraphics[width=\textwidth]{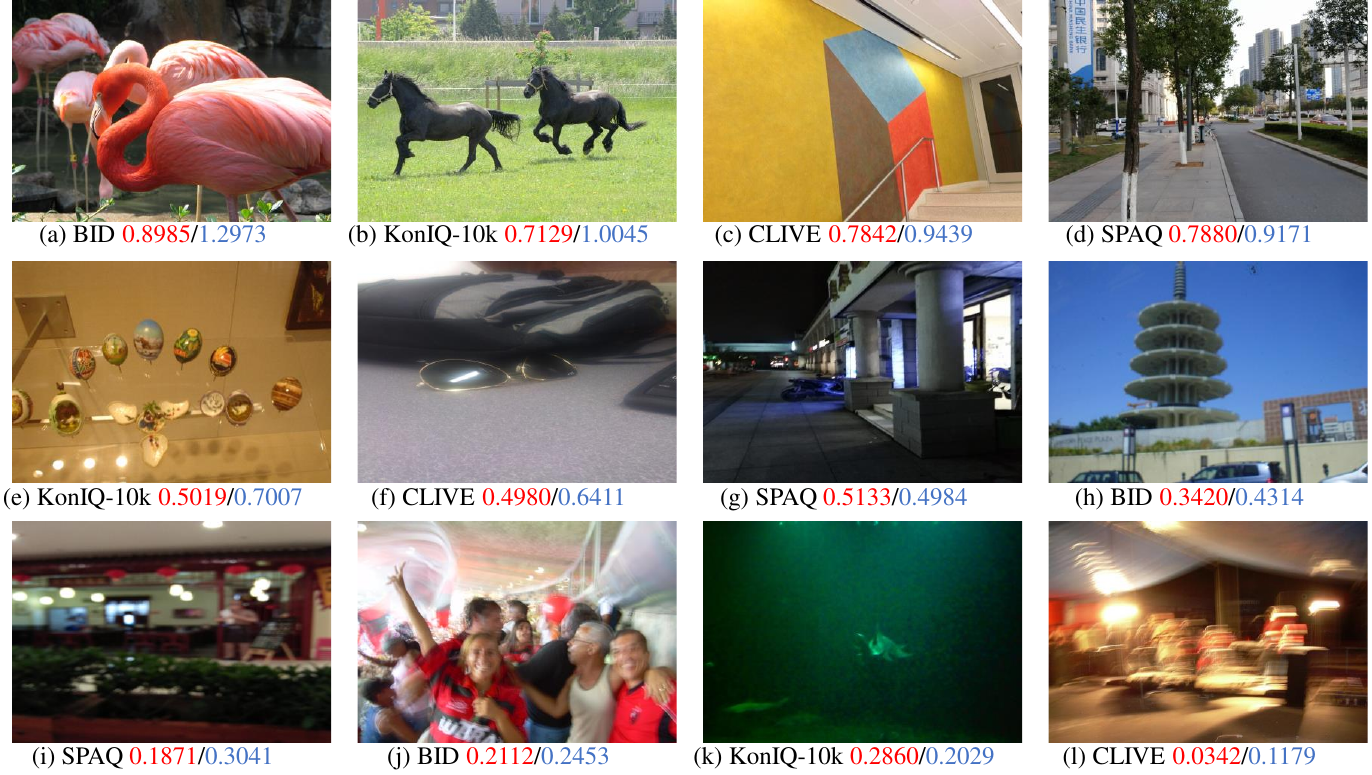}
    \caption{Visual examples from different databases aligned in the predicted quality scores generated by UNQA. (a)-(d) Images with good predicted quality. (e)-(h) Images with fair predicted quality. (i)-(l) Images with poor predicted quality. \textcolor{red}{Red}/\textcolor{blue}{Blue} denotes the \textcolor{red}{MOSs}/\textcolor{blue}{predict scores} for each image. In each row, images are arranged from left to right in descending order of predicted quality.}
\label{fig:visual}
\end{figure*}
\subsection{Experimental Settings}
\textbf{Databases}. We conduct experiments on the three AQA databases, four IQA databases, three VQA databases, and two AVQA databases. Table \ref{tab:AQA Database} summarizes the main information of the three AQA databases (ITU-T \cite{recommendation1998itu}, NOIZEUS \cite{hu2006subjective}, and TCDVoIP \cite{harte2015tcd}) and four IQA databases (BID \cite{ciancio2010no}, CLIVE \cite{ghadiyaram2015massive}, KonIQ10K \cite{hosu2020koniq}, and SPAQ \cite{fang2020perceptual}). Table \ref{tab:VQA Database} summarizes the main information of three VQA databases and two AVQA databases. YouTube-UGC \cite{wang2019youtube}, KoNViD-1k \cite{hosu2017konstanz}, and LIVE-VQC \cite{sinno2018large} are the VQA databases, while SJTU-UAV \cite{cao2023UGC} and LIVE-SJTU \cite{min2020study} are the AVQA databases.

\textbf{Compared Methods}. We compare our proposed model with the SOTA NR AQA, IQA, VQA, and AVQA models, including:
\begin{itemize}
    \item Deep learning-based AQA models: MOSNet \cite{lo2019mosnet}, Wenets \cite{catellier2019wenets}, Quality-Net \cite{fu2018quality}, STOI-Net \cite{zezario2020stoi}, NRMusic \cite{li2013non}, and ARN-50 \cite{cao2023subjective}.
    \item Handcrafted feature-based IQA models: NIQE \cite{mittal2012making} and BRISQUE \cite{6272356}.
    \item Deep learning-based IQA models:  CNNIQA \cite{kang2014convolutional}, DB-CNN \cite{zhang2020blind}, HyperIQA \cite{su2020blindly}, and StairIQA \cite{sun2023blind}.
    \item Handcrafted feature-based VQA models: V-BLIINDS \cite{saad2014blind}, VIDEVAL \cite{tu2021ugc}, and RAPIQUE \cite{tu2021rapique}.
    \item Deep learning-based VQA models: VSFA \cite{Li2019}, SimpleVQA \cite{sun2022deep} and FastVQA \cite{wu2022fast}.
    \item Handcrafted feature-based AVQA models: three BB-SESQA models \cite{Martinez2018} with different fusing methods (Linear, Power and Minkowski).
    \item Deep learning-based AVQA models: No-Reference Deep Neural Families of A/V Quality Predictors (NR-DNFAVQ) \cite{min2020study}, DNN-RNT, and DNN-SND \cite{cao2021deep}.
\end{itemize}
The above methods address the single modality QA problems and train the models on the single QA databases except StairIQA. StairIQA simultaneously trains on four IQA databases and obtains four separate model parameters to test on each IQA database. For ARN-50, we first utilize the short-time Fourier transform (STFT) to calculate the spectrogram of each audio segment. The spectrograms are fed into ResNet-50 to extract spectrogram features which are subsequently averaged over all audio segments to produce the audio features. Then, feed spectrogram features into SVR to predict audio quality scores.

\textbf{Evaluation Criteria}. We use Spearman’s rank-order correlation coefficient (SROCC) and Pearson’s linear correlation coefficient (PLCC) to evaluate the effectiveness of AVQA methods. SRCC measures the prediction monotonicity, while PLCC measures the prediction accuracy. Better AVQA methods should have larger SRCC and PLCC values.

\textbf{Details}. We utilize the ConvNext V2 \cite{woo2023convnext} pre-trained on ImageNet as the backbone of the spatial feature extraction module and the SlowFast R50 \cite{feichtenhofer2019slowfast} pre-trained on the Kinetics 400 database \cite{kay2017kinetics} as the motion feature extraction module. During the overall joint training, the weights of the SlowFast R50 are fixed. For the spatial feature extraction module, we resize the resolution of the minimum dimension of inputs as $520$ while maintaining their aspect ratio, and then crop inputs with the resolution of $384\times384$ at the center. For the motion feature extraction module, the resolution of the video frame is resized to $224\times224$. The learning rate is set to $0.00001$. We train $20$, $10$, and $10$ epochs in the three steps of our multi-modality training strategy, respectively. We randomly split each database into a training set, a validation set, and a testing set at $7:1:2$. In each step of our multi-modality training strategy, we train models only on the training set and find the top models on the validation set. We test the top models of the regressor refinement step on the testing set as the final results. This procedure is randomly repeated $10$ times to prevent performance bias, and mean results are reported for comparison. Our proposed model is implemented with PyTorch.

\subsection{Comparison with the SOTA}
Since there are no other QA models jointly trained across multiple modalities, we compare UNQA with 24 SOTA QA methods trained on a specific database, as shown in Table \ref{tab:all}. UNQA is jointly trained on all QA databases to generate a comprehensive model through our multi-modality strategy. StairIQA \cite{sun2023blind} utilizes an iterative mixed database training (IMDT) strategy to simultaneously train the model on four IQA databases, with different model parameters for each database. While UNQA-Single and the other QA methods are trained separately on each database to obtain database-specific models. From the results, we have three interesting observations: Firstly, our proposed model UNQA generally outperforms SOTA QA methods. It verifies that our model is capable of learning unified feature representations on multiple QA databases and predicting the qualities across audio, image, video, and A/V contents via the same model parameters. Secondly, the databases with small sizes, such as BID, CLIVE, LIVE-VQC, and SJTU-UAV, benefit more from our model and our multi-modality training strategy. StairIQA simultaneously trains on four IQA databases, which also achieves good performances on the BID and CLIVE databases. It means that our model can learn more general feature representations from multiple databases to improve the performance on the small size databases. Finally, UNQA achieves relatively better performance than UNQA-Single, which demonstrates that our multi-modality training strategy can help our model learn more general feature representations and improve performance on each QA database.

\textbf{Qualitative Results.} We conduct a qualitative analysis of UNQA by sampling images across four IQA databases, as shown in Fig. \ref{fig:visual}. Our proposed multi-modality training strategy differs from pair-wise learning methods, which do not generate pairs of images from two different databases. It is capable of aligning MOSs from different databases in a perceptually meaningful way.

\textbf{Parameters Parameters.} Since most of deep learning-based QA models have better performance than the handcrafted feature-based QA models, we compare the total number of parameters of UNQA and deep learning-based QA models in Table \ref{tab:params}. Except UNQA, the other deep learning-based QA models are database-specific models, which are trained separately on the corresponding databases to obtain multiple trained models. We report the total number of parameters across all trained models in the ``Total params'' row of Table \ref{tab:params}. Although CNNIQA owns the least number of parameters, it performs poorly on the IQA databases. Our proposed model UNQA has a relatively small number of parameters. And the parameters of UNQA will not increase with the increase of training databases. 
\begin{table*}[!htb]
    \centering
    \caption{Ablation study on the multi-modality training strategy. The best values are marked in boldface.}
    \label{tab:ablation1}
    \begin{spacing}{1.19}
    \resizebox{\textwidth}{!}{
    \huge
    \begin{tabular}{cc||cccc|ccc|ccc|cc}
    \toprule
    Criteria & Method & BID & CLIVE & KonIQ-10k & SPAQ & YouTube-UGC & KoNViD-1k & LIVE-VQC & ITU & NOIZEUS & TCDVoIP & SJTU-UAV & LIVE-SJTU\\
    \midrule
    \multirow{4}{*}{SRCC} & LRS & 0.8168 & 0.7886 & 0.7940 & 0.8813 & 0.7217 & 0.8307 & 0.7912 & 0.5696 & 0.6872 & 0.5672 & 0.7493 & 0.8977\\
    ~ & MDT & 0.8743 & 0.8635 & 0.8898 & 0.9098 & 0.8124 & 0.8429 & 0.8074 & 0.6030 & 0.8131 & 0.6211 & 0.8031 & 0.9293\\
    ~ & WTS & 0.8789 & 0.8663 & 0.8914 & 0.9101 & 0.8386 & 0.8624 & 0.8125 & 0.7078 & 0.8363 & 0.6152 & 0.8234 & 0.9494\\
    ~ & UNQA & \textbf{0.8812} & \textbf{0.8744} & \textbf{0.8930} & \textbf{0.9127} & \textbf{0.8411} & \textbf{0.8666} & \textbf{0.8235} & \textbf{0.8565} & \textbf{0.9225} & \textbf{0.8416} & \textbf{0.8301} & \textbf{0.9541}\\
    \midrule
    \multirow{4}{*}{PLCC} & LRS & 0.8480 & 0.8264 & 0.8314 & 0.8811 & 0.7315 & 0.8286 & 0.8056 & 0.5641 & 0.6621 & 0.6024 & 0.7619 & 0.8793\\
    ~ & MDT & 0.9019 & 0.8905 & 0.9113 & 0.9136 & 0.8140 & 0.8441 & 0.8224 & 0.6160 & 0.8076 & 0.6327 & 0.8068 & 0.9243\\
    ~ & WTS & 0.9118 & 0.8946 & 0.9148 & 0.9134 & 0.8433 & 0.8649 & 0.8345 & 0.7203 & 0.8293 & 0.6656 & 0.8348 & 0.9274\\
    ~ & UNQA & \textbf{0.9140} & \textbf{0.9027} & \textbf{0.9153} & \textbf{0.9174} & \textbf{0.8457} & \textbf{0.8669} & \textbf{0.8351} & \textbf{0.8662} & \textbf{0.9269} & \textbf{0.8719} & \textbf{0.8379} & \textbf{0.9309}\\
    \bottomrule
    \end{tabular}
    }
    \end{spacing}
\end{table*}
\begin{table*}[!htb]
    \centering
    \caption{Ablation study of three steps in the multi-modality training strategy. FS, SS, and TS represent the first step, the second step, and the third step of our multi-modality training strategy, respectively. The best values are marked in boldface.}
    \label{tab:withoutstep}
    \begin{spacing}{1.19}
    \resizebox{\textwidth}{!}{
    \huge
    \begin{tabular}{cc||cccc|ccc|ccc|cc}
    \toprule
    Criteria & Method & BID & CLIVE & KonIQ-10k & SPAQ & YouTube-UGC & KoNViD-1k & LIVE-VQC & ITU & NOIZEUS & TCDVoIP & SJTU-UAV & LIVE-SJTU\\
    \midrule
    \multirow{4}{*}{SRCC} & \emph{w/o} FS & 0.8724 & 0.8720 & 0.8819 & 0.9094 & 0.8165 & 0.8612 & 0.8188 & 0.6179 & 0.8128 & 0.6309 & 0.8098 & 0.9016\\
    ~ & \emph{w/o} SS & 0.8736 & 0.8647 & 0.8844 & 0.9105 & 0.8285 & 0.8650 & 0.8098 & 0.3134 & 0.3956 & 0.5193 & 0.7969 & 0.9363\\
    ~ & \emph{w/o} TS & 0.8787 & 0.8709 & 0.8882 & 0.9117 & 0.8337 & 0.8573 & 0.8203 & 0.8516 & 0.9221 & 0.8411 & 0.8195 & 0.9514\\
    ~ & UNQA & \textbf{0.8812} & \textbf{0.8744} & \textbf{0.8930} & \textbf{0.9127} & \textbf{0.8411} & \textbf{0.8666} & \textbf{0.8235} & \textbf{0.8565} & \textbf{0.9225} & \textbf{0.8416} & \textbf{0.8301} & \textbf{0.9541}\\
    \midrule
    \multirow{4}{*}{PLCC} & \emph{w/o} FS & 0.9114 & 0.8955 & 0.9071 & 0.9134 & 0.8214 & 0.8586 & 0.8313 & 0.6338 & 0.8084 & 0.6537 & 0.8202 & 0.8842\\
    ~ & \emph{w/o} SS & 0.9086 & 0.8958 & 0.9007 & 0.9153 & 0.8359 & 0.8610 & 0.8239 & 0.3074 & 0.3973 & 0.5336 & 0.8002 & 0.9121\\
    ~ & \emph{w/o} TS & 0.9125 & 0.9001 & 0.9112 & 0.9162 & 0.8380 & 0.8595 & 0.8293 & 0.8621 & 0.9260 & 0.8709 & 0.8264 & 0.9309\\
    ~ & UNQA & \textbf{0.9140} & \textbf{0.9027} & \textbf{0.9153} & \textbf{0.9174} & \textbf{0.8457} & \textbf{0.8669} & \textbf{0.8351} & \textbf{0.8662} & \textbf{0.9269} & \textbf{0.8719} & \textbf{0.8379} & \textbf{0.9309}\\
    \bottomrule
    \end{tabular}
    }
    \end{spacing}
\end{table*}
\begin{table*}[!htb]
    \centering
    \caption{Ablation study on the feature extraction modules of UNQA. S, M, and A indicate the spatial feature extraction, motion feature extraction, and audio feature extraction modules, respectively. The best values are marked in boldface.}
    \label{tab:ablation3}
    \begin{spacing}{1.19}
    \resizebox{\textwidth}{!}{
    \huge
    \begin{tabular}{cc||cccc|ccc|ccc|cc}
    \toprule
    Criteria & Method & BID & CLIVE & KonIQ-10k & SPAQ & YouTube-UGC & KoNViD-1k & LIVE-VQC & ITU & NOIZEUS & TCDVoIP & SJTU-UAV & LIVE-SJTU\\
    \midrule
    \multirow{4}{*}{SRCC} & \emph{w/o} S & - & - & - & - & 0.4707 & 0.7495 & 0.7084 & 0.6938 & 0.9134 & 0.7346 & 0.5483 & 0.8749\\
    ~ & \emph{w/o} M & 0.8693 & 0.8646 & 0.8781 & 0.9110 & 0.8056 & 0.8595 & 0.7843 & 0.7668 & 0.8986 & 0.7516 & 0.8031 & 0.9537\\
    ~ & \emph{w/o} A & 0.8723 & 0.8676 & 0.8806 & 0.9101 & 0.8395 & 0.8534 & 0.8077 & - & - & - & 0.8011 & 0.7369\\
    ~ & UNQA & \textbf{0.8812} & \textbf{0.8744} & \textbf{0.8930} & \textbf{0.9127} & \textbf{0.8411} & \textbf{0.8666} & \textbf{0.8235} & \textbf{0.8565} & \textbf{0.9225} & \textbf{0.8416} & \textbf{0.8301} & \textbf{0.9541}\\
    \midrule
    \multirow{4}{*}{PLCC} & \emph{w/o} S & - & - & - & - & 0.5237 & 0.7352 & 0.7283 & 0.7109 & 0.9133 & 0.7557 & 0.5487 & 0.8430\\
    ~ & \emph{w/o} M & 0.8988 & 0.8931 & 0.9037 & 0.9144 & 0.8228 & 0.8595 & 0.8041 & 0.7877 & 0.8960 & 0.8061 & 0.8151 & 0.9241\\
    ~ & \emph{w/o} A & 0.9078 & 0.8979 & 0.9106 & 0.9159 & 0.8397 & 0.8576 & 0.8197 & - & - & - & 0.8128 & 0.7206\\
    ~ & UNQA & \textbf{0.9140} & \textbf{0.9027} & \textbf{0.9153} & \textbf{0.9174} & \textbf{0.8457} & \textbf{0.8669} & \textbf{0.8351} & \textbf{0.8662} & \textbf{0.9269} & \textbf{0.8719} & \textbf{0.8379} & \textbf{0.9309}\\
    \bottomrule
    \end{tabular}
    }
    \end{spacing}
\end{table*}

\subsection{Ablation Study}
We conduct some ablation studies to verify the rationality of our multi-modality training strategy, the feature extraction modules, and the regression module of UNQA. We also analyze the effect of each modality during joint training.

\textbf{Multi-Modality Training Strategy.} 
To verify the effectiveness of our proposed multi-modality training strategy, we compare it with different training strategies. For linear re-scaling (\textbf{LRS}), we train UNQA as a standard regression model. We linearly rescale the subjective scores of each QA database to $[0, 1]$ and jointly train UNQA on all databases with MSE. The mixed database training (\textbf{MDT}) strategy \cite{li2021unified} is proposed for joint training on multiple VQA databases, which consists of three stages: relative quality assessor, nonlinear mapping, and database-specific perceptual scale alignment. We also utilize MDT to jointly train our model on all databases with the sum of the monotonicity-induced loss, linearity-induced loss, and error-induced loss. To verify the effectiveness of the adaptive weighted task sampling method, we utilize our proposed multi-modality training strategy with the weighted task sampling (\textbf{WTS}) to jointly train our model. The results are listed in Table \ref{tab:ablation1}. It can be seen that our proposed multi-modality training strategy achieves the best performance on all databases. The total number of training steps in our proposed strategy does not exceed the sum of all single-database training steps. Our proposed multi-modality training strategy is efficient and practical. Adaptive weighted task sampling can make different modalities more harmonious during joint training.

\textbf{Steps in Multi-Modality Training Strategy.}
In order to further verify the rationality of our multi-modality training strategy, we remove the first step (\textbf{\emph{w/o} FS}), the second step (\textbf{\emph{w/o} SS}), and the third step (\textbf{\emph{w/o} TS}) of our multi-modality training strategy to train UNQA, respectively. As shown in Table \ref{tab:withoutstep}, removing any step in our multi-modality training strategy leads to performance decline. It verifies the effectiveness of three steps in our multi-modality training strategy.

\textbf{Feature Extraction Modules.}
We test UNQA without the spatial feature extraction module (\textbf{\emph{w/o} S}), without the motion feature extraction module (\textbf{\emph{w/o} M}), and without the audio feature extraction module (\textbf{\emph{w/o} A}) to investigate the effect of these three kinds of features, respectively. The results are listed in Table \ref{tab:ablation3}. It can be observed that \emph{w/o} S is inferior to \emph{w/o} M and \emph{w/o} A, and all of them are inferior to the proposed model, which indicates that the three feature extraction modules are all beneficial to UNQA and the spatial feature extraction module is extremely important. Then, we further do some ablation studies for the spatial feature extraction module. We begin with a \textbf{baseline} that utilizes the pre-trained ResNet-50 on ImageNet as the spatial feature extractor. We then test UNQA without MHSA (\textbf{\emph{w/o} MHSA}). As shown in Table \ref{tab:ablation2}, UNQA utilizing the ConvNext V2 with MHSA to extract the spatial feature leads to performance improvement compared to the baseline. 
\begin{table*}[!htb]
    \centering
    \caption{Ablation study on the spatial feature extraction module. The best values are marked in boldface.}
    \label{tab:ablation2}
    \begin{spacing}{1.19}
    \resizebox{\textwidth}{!}{
    \huge
    \begin{tabular}{cc||cccc|ccc|ccc|cc}
    \toprule
    Criteria & Method & BID & CLIVE & KonIQ-10k & SPAQ & YouTube-UGC & KoNViD-1k & LIVE-VQC & ITU & NOIZEUS & TCDVoIP & SJTU-UAV & LIVE-SJTU\\
    \midrule
    \multirow{3}{*}{SRCC} & Basline & 0.8602 & 0.8584 & 0.8754 & 0.9033 & 0.8144 & 0.8567 & 0.8023 & \textbf{0.8784} & 0.9110 & 0.8417 & 0.8132 & 0.9283\\
    ~ & \emph{w/o} MHSA & 0.8612 & 0.8550 & 0.8860 & 0.9089 & 0.8337 & 0.8647 & 0.8182 & 0.8401 & 0.9134 & 0.8425 & 0.8191 & 0.9373\\
    ~ & UNQA & \textbf{0.8812} & \textbf{0.8744} & \textbf{0.8930} & \textbf{0.9127} & \textbf{0.8411} & \textbf{0.8666} & \textbf{0.8235} & 0.8565 & \textbf{0.9225} & \textbf{0.8416} & \textbf{0.8301} & \textbf{0.9541}\\
    \midrule
    \multirow{3}{*}{PLCC} & Basline & 0.8943 & 0.8879 & 0.8981 & 0.9083 & 0.7965 & 0.8521 & 0.8118 & \textbf{0.8845} & 0.9127 & 0.8710 & 0.8025 & 0.9098\\
    ~ & \emph{w/o} MHSA & 0.8995 & 0.8871 & 0.9076 & 0.9132 & 0.8361 & 0.8606 & 0.8272 & 0.8505 & 0.9147 & 0.8704 & 0.8290 & 0.9163\\
    ~ & UNQA & \textbf{0.9140} & \textbf{0.9027} & \textbf{0.9153} & \textbf{0.9174} & \textbf{0.8457} & \textbf{0.8669} & \textbf{0.8351} & 0.8662 & \textbf{0.9269} & \textbf{0.8719} & \textbf{0.8379} & \textbf{0.9309}\\
    \bottomrule
    \end{tabular}
    }
    \end{spacing}
\end{table*}
\begin{table*}[!htb]
    \centering
    \caption{Ablation study on the regression module. The best values are marked in boldface.}
    \label{tab:regression}
    \begin{spacing}{1.19}
    \resizebox{\textwidth}{!}{
    \huge
    \begin{tabular}{cc||cccc|ccc|ccc|cc}
    \toprule
    Criteria & Method & BID & CLIVE & KonIQ-10k & SPAQ & YouTube-UGC & KoNViD-1k & LIVE-VQC & ITU & NOIZEUS & TCDVoIP & SJTU-UAV & LIVE-SJTU\\
    \midrule
    \multirow{3}{*}{SRCC} & MLP & 0.8640 & 0.8515 & 0.8725 & 0.9006 & 0.8289 & 0.8435 & 0.7961 & 0.8748 & 0.9259 & 0.8405 & 0.7766 & 0.8810\\
    ~ & Transformer & 0.8773 & 0.8701 & 0.8834 & 0.9081 & 0.8317 & 0.8542 & 0.8065 & \textbf{0.8833} & \textbf{0.9305} & 0.8408 & 0.7991 & 0.9202\\
    ~ & UNQA & \textbf{0.8812} & \textbf{0.8744} & \textbf{0.8930} & \textbf{0.9127} & \textbf{0.8411} & \textbf{0.8666} & \textbf{0.8235} & 0.8565 & 0.9225 & \textbf{0.8416} & \textbf{0.8301} & \textbf{0.9541}\\
    \midrule
    \multirow{3}{*}{PLCC} & MLP & 0.8943 & 0.8811 & 0.8968 & 0.9049 & 0.8320 & 0.8602 & 0.7958 & 0.8820 & 0.9299 & 0.8669 & 0.7903 & 0.8506\\
    ~ & Transformer & 0.9106 & 0.8986 & 0.9068 & 0.9132 & 0.8388 & 0.8551 & 0.8266 & \textbf{0.8899} & \textbf{0.9355} & \textbf{0.8728} & 0.8108 & 0.8908\\
    ~ & UNQA & \textbf{0.9140} & \textbf{0.9027} & \textbf{0.9153} & \textbf{0.9174} & \textbf{0.8457} & \textbf{0.8669} & \textbf{0.8351} & 0.8662 & 0.9269 & 0.8719 & \textbf{0.8379} & \textbf{0.9309}\\
    \bottomrule
    \end{tabular}
    }
    \end{spacing}
\end{table*}
\begin{table*}[!htb]
    \centering
    \caption{Ablation study of the effect of each modality. The best values are marked in boldface.}
    \label{tab:modality}
    \begin{spacing}{1.19}
    \resizebox{\textwidth}{!}{
    \huge
    \begin{tabular}{cc||cccc|ccc|ccc|cc}
    \toprule
    Criteria & Method & BID & CLIVE & KonIQ-10k & SPAQ & YouTube-UGC & KoNViD-1k & LIVE-VQC & ITU & NOIZEUS & TCDVoIP & SJTU-UAV & LIVE-SJTU\\
    \midrule
    \multirow{5}{*}{SRCC} & \emph{w/o} IQA & - & - & - & - & 0.8264 & 0.8552 & 0.8030 & 0.8006 & 0.9092 & 0.8149 & 0.8290 & 0.9526\\
    ~ & \emph{w/o} VQA & \textbf{0.8844} & 0.8802 & \textbf{0.9040} & \textbf{0.9161} & - & - & - & 0.8232 & 0.9038 & 0.8164 & 0.8139 & 0.9393\\
    ~ & \emph{w/o} AQA & 0.8745 & 0.8692 & 0.8903 & 0.9111 & 0.8400 & 0.8581 & 0.8095 & - & - & - & 0.8193 & 0.9413\\
    ~ & \emph{w/o} AVQA & 0.8801 & \textbf{0.8809} & 0.8961 & 0.9131 & 0.8407 & 0.8641 & 0.8166 & 0.8492 & 0.9034 & 0.8221 & - & -\\
    ~ & UNQA & 0.8812 & 0.8744 & 0.8930 & 0.9127 & \textbf{0.8411} & \textbf{0.8666} & \textbf{0.8235} & \textbf{0.8565} & \textbf{0.9225} & \textbf{0.8416} & \textbf{0.8301} & \textbf{0.9541}\\
    \midrule
    \multirow{5}{*}{PLCC} & \emph{w/o} IQA & - & - & - & - & 0.8319 & 0.8552 & 0.8145 & 0.8164 & 0.9099 & 0.8484 & 0.8335 & 0.9279\\
    ~ & \emph{w/o} VQA & \textbf{0.9186} & 0.9074 & 0.9128 & 0.9210 & - & - & - & 0.8347 & 0.9050 & 0.8500 & 0.8219 & 0.9174\\
    ~ & \emph{w/o} AQA & 0.9106 & 0.8988 & 0.9114 & 0.9156 & 0.8404 & 0.8605 & 0.8219 & - & - & - & 0.8306 & 0.9239\\
    ~ & \emph{w/o} AVQA & 0.9150 & \textbf{0.9078} & \textbf{0.9163} & \textbf{0.9178} & 0.8430 & 0.8665 & 0.8291 & 0.8583 & 0.9028 & 0.8532 & - & -\\
    ~ & UNQA & 0.9140 & 0.9027 & 0.9153 & 0.9174 & \textbf{0.8457} & \textbf{0.8669} & \textbf{0.8351} & \textbf{0.8662} & \textbf{0.9269} & \textbf{0.8719} & \textbf{0.8379} & \textbf{0.9309}\\
    \bottomrule
    \end{tabular}
    }
    \end{spacing}
\end{table*}

\textbf{Regression Module.}
Since the regression module of UNQA consists of the four modality-specific regression modules, we try to utilize a single regression module instead of the four modality-specific regression modules. We utilize the single MLP and transformer as the regression module, respectively. As shown in Table \ref{tab:regression}, UNQA performs better than utilizing the single regression module. Although Transformer has a lower performance than UNQA, Transformer still achieves the promising performance and unifies the regression module when facing different modality inputs.

\textbf{Effect of Each Modality.} Three feature extraction modules work together across different modalities. In Table \ref{tab:modality}, joint training with IQA databases has a positive effect on the performances of VQA and AVQA databases, while joint training with VQA and AVQA databases slightly declines the performance on IQA databases. It is because with VQA databases, the model can learn a combined feature representation of motion and spatial features, and when used for the image modality, the absence of motion features will degrade the model performance to a certain extent.
\begin{table}[!t]
    \centering
    \caption{The basic information of the seven testing databases. FR: Framerate.}
    \begin{spacing}{1.1}
    \resizebox{0.5\textwidth}{!}{
    \begin{tabular}{c|c| ccccc}
    \toprule
    \multirow{3}{*}{IQA} & Attribute & Number & Resolution & Distortion\\
    \cline{2-7}
    ~ & CID2013 \cite{virtanen2014cid2013} & 474 & 1600$\times$1200 & Authentic\\
    ~ & LIVE \cite{sheikh2006statistical} & 779 & $\sim$768$\times$512 & Synthetic\\
    \bottomrule
    \multirow{3}{*}{VQA} & Attribute & Number & Resolution & Distortion & Length & FR\\
    \cline{2-7}
    ~ & CVD2014 \cite{nuutinen2016cvd2014}& 234 & 480p,720p & Authentic & 10-25s & 10-25\\
    ~ & LIVE-Qualcomm \cite{ghadiyaram2017capture} & 208 & 1080p & Authentic & 15s & 30 \\
    \midrule
    \multirow{3}{*}{AQA} & Attribute & Number & \multicolumn{4}{c}{Distortion}\\
    \cline{2-7}
    ~ & NISQA\_LIVETALK \cite{mittag2021nisqa} & 232 & \multicolumn{4}{c}{Real phone and VoIP calls distortion}\\
    ~ & NISQA\_P501 \cite{mittag2021nisqa} & 240 & \multicolumn{4}{c}{Simulated distortions and live VoIP calls}\\
    \midrule
    \multirow{2}{*}{AVQA} & Attribute & Number & Resolution & Distortion & Length & FR\\
    \cline{2-7}
    ~ & UnB-AVC \cite{martinez2014no}& 72 & 720p & Synthetic & 8s & 30\\
    \bottomrule
    \end{tabular}
    }
    \end{spacing}
    \label{tab:test Database}
\end{table}
\begin{table*}[!htb]
    \centering
    \caption{The cross-database evaluation of UNQA and SOTA QA methods on IQA, VQA, AQA, and AVQA databases. StairIQA is simultaneously trained on four IQA databases. The other IQA, VQA, AQA, and AVQA methods are trained on SPAQ, KoNViD-1k, TCDVoIP, and UnB-AVC database, respectively. The best and second-best values for each metric are marked in boldface and underlined, respectively.}
    \label{tab:VQA-corss}
    \begin{spacing}{1.1}
    \resizebox{\textwidth}{!}{
    \begin{tabular}{c||cc cc ||c || cc cc||c||cc cc||c||cc}
    \toprule
    \multirow{2}{*}{IQA Method} & \multicolumn{2}{c}{CID2013} & \multicolumn{2}{c||}{LIVE} & \multirow{2}{*}{VQA Method} & \multicolumn{2}{c}{CVD2014} & \multicolumn{2}{c||}{LIVE-Qualcomm} & \multirow{2}{*}{AQA Method} & \multicolumn{2}{c}{NISQA\_LIVETALK} & \multicolumn{2}{c||}{NISQA\_P501} & \multirow{2}{*}{AVQA Method} & \multicolumn{2}{c}{UnB-AVC}\\
    ~  & SRCC & PLCC & SRCC & PLCC & ~ & SRCC & PLCC & SRCC & PLCC & ~ & SRCC & PLCC & SRCC & PLCC & ~ & SRCC & PLCC\\
    \midrule
    BRISQUE & 0.4017 & 0.4171 & 0.4444 & 0.4599 & VIDEVAL & 0.5261 & 0.5076 & 0.4748 & 0.5007 & Wenets & 0.1704 & 0.1214 & 0.3031 & 0.3431 & Linear & 0.8594 & 0.7708\\
    CNNIQA & 0.6091 & 0.6064 & 0.1604 & -0.2769 & RAPIQUE & 0.5478 & 0.6328 & 0.1134 & 0.1529 & Quality-Net & 0.2534 & 0.1632 & 0.3219 & 0.3451 & Minkowski & \underline{0.8671} & 0.7105 \\
    DBCNN & 0.6342 & 0.6723 & 0.7623 & 0.7271 & VSFA & 0.6144 & 0.6746 & 0.5973 & 0.6389 & STOI-Net & 0.1221	& 0.2034 & 0.2053 & 0.3164 & DNFAVQ & 0.8406 & 0.8344\\
    HyerIQA & 0.6301 & 0.6588 & 0.2413 & 0.1898 & SimpleVQA & 0.7012 & 0.7485 & 0.6077 & 0.6287 & NRMusic & 0.2855 & 0.1852 & 0.3577 & 0.3541 & DNN-RNT & 0.7871 & 0.7743\\
    StairIQA & \underline{0.6990} & \underline{0.7489} & \underline{0.7760} & \underline{0.7588} & Fast-VQA & \underline{0.7075} & \underline{0.7539} & \underline{0.6561} & \underline{0.6787} & ARN-50 & \underline{0.4406} & \underline{0.4429} & \underline{0.4991} & \underline{0.4967} & DNN-SND & 0.8213 & \underline{0.8626}\\
    \midrule
    UNQA & \textbf{0.7962} & \textbf{0.8037} & \textbf{0.8901} & \textbf{0.8791} & UNQA & \textbf{0.7854} & \textbf{0.8055} & \textbf{0.6672} & \textbf{0.6897} & UNQA & \textbf{0.4933} & \textbf{0.4933} & \textbf{0.5775} & \textbf{0.5775} & UNQA & \textbf{0.9029} & \textbf{0.8634}\\
    \bottomrule
    \end{tabular}
    }
    \end{spacing}
\end{table*}
\begin{figure*}[!t]
 \centering
    \includegraphics[width=\textwidth]{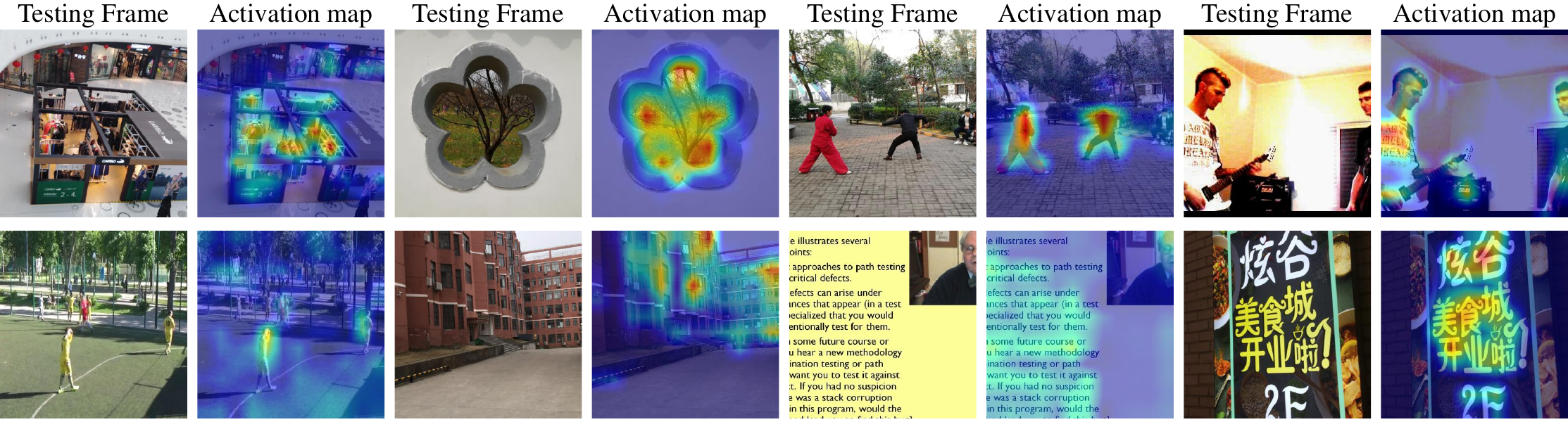}
    \caption{Sample frames of eight testing videos and the corresponding activation maps obtained via the Grad-CAM method \cite{selvaraju2017grad}.}
\label{fig:GradCam}
\end{figure*}
\subsection{Cross-Database Evaluation}
The generalization ability is important for the QA models to predict the quality scores of unseen distortions and content. Here, we conduct the cross-database evaluation to verify the generalization ability of UNQA. Since UNQA is jointly trained on the $12$ QA databases, we select other seven publicly available QA databases to test the performance of cross-database evaluation, including two IQA databases, two VQA databases, two AQA databases, and one AVQA database. The basic information of the above seven testing databases is shown in Table \ref{tab:test Database}. 

Since there are no other QA models jointly trained across multiple modalities, we compare the generalization ability of UNQA with 20 SOTA QA methods. UNQA is jointly trained on the $12$ QA databases which include four IQA databases (BID \cite{ciancio2010no}, CLIVE \cite{ghadiyaram2015massive}, KonIQ10K \cite{hosu2020koniq}, SPAQ \cite{fang2020perceptual}), three VQA databases (YouTube-UGC \cite{wang2019youtube}, KoNViD-1k \cite{hosu2017konstanz}, LIVE-VQC \cite{sinno2018large}), three AQA databases (ITU-T \cite{recommendation1998itu}, NOIZEUS \cite{hu2006subjective}, and TCDVoIP \cite{harte2015tcd}) and two AVQA databases (SJTU-UAV \cite{cao2023UGC} and LIVE-SJTU \cite{min2020study}). StairIQA is simultaneously trained on four IQA databases. The other IQA, VQA, AQA, and AVQA methods are trained on SPAQ, KoNViD-1k, TCDVoIP, and UnB-AVC database, respectively. The experimental results are listed in Table \ref{tab:VQA-corss}, which verifies the favorable generalizability of UNQA. It demonstrates that UNQA jointly trained on multiple modalities databases does learn general feature representations and is robust to assess the quality of audio, image, video, and A/V content sampled from various distributions. 

\begin{figure}[!tb]
\captionsetup[subfigure]{justification=centering}
\centering
  \includegraphics[width=0.95\linewidth]{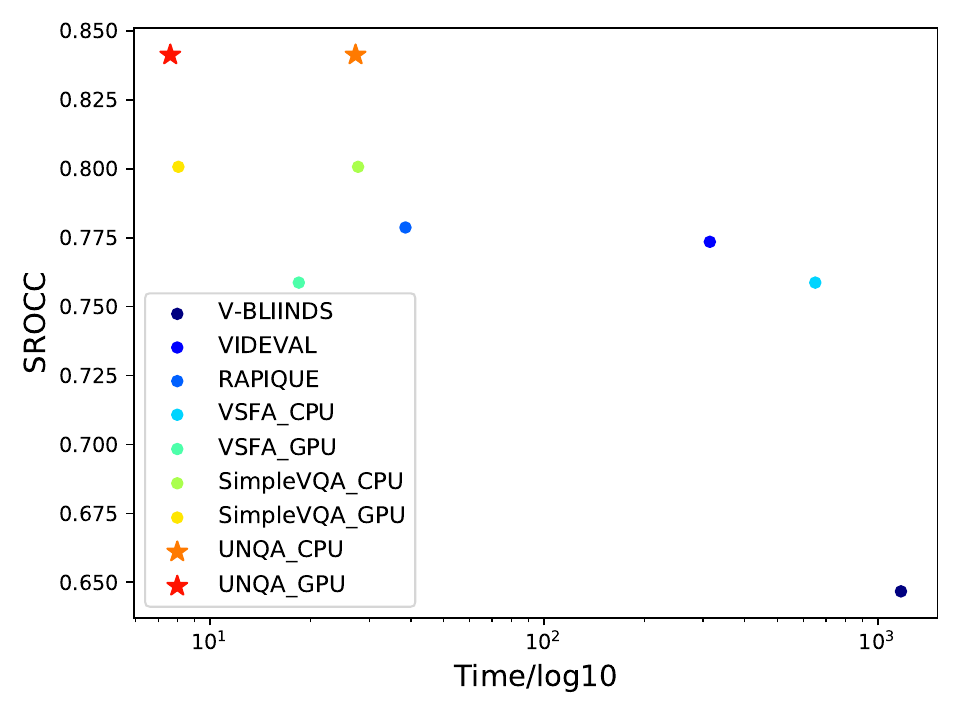}
  \caption{Scatter plots of the SRCC results (collected from Table \ref{tab:all}) versus the running time of each QA method.}
\label{fig:time}
\end{figure}
\subsection{Visual Analysis}
We utilize the Grad-CAM method \cite{selvaraju2017grad} to highlight crucial regions with the greatest influence on the quality prediction of UNQA. It can provide interpretability for UNQA and reveal the specific image or frame areas our model focuses on during quality prediction. Fig. \ref{fig:GradCam} depicts the sample frames of eight testing video frames and the corresponding activation maps obtained via the commonly used Grad-CAM method. It can be observed that UNQA focuses on visually salient areas to predict quality scores, such as people, text, buildings, and so on. 

\subsection{Computational Complexity}
Since computational efficiency is very important in practical applications, we compare the computational complexity in this section. All models are tested on a computer with Intel Core i7-4790K CPU@4.00GHz, 24GB RAM, and NVIDIA GeForce GTX TITAN X. The handcrafted feature-based models are tested using CPU. The deep learning-based models are tested using GPU and CPU, respectively. We report the running time for a 25fps video with the resolution of 1080p and time duration of 8s in Fig. \ref{fig:time}. It can be observed that UNQA has the lowest running time compared with other models. The very low computational complexity makes UNQA suitable for practical applications.

\section{Conclusion}
In this paper, we have first proposed a Unified No-reference Quality Assessment model (UNQA) which consists of three feature extraction modules and four modality-specific regression modules. It is the first time that utilizes a unified model to predict quality scores of audio, image, video, and A/V content, simultaneously. Secondly, we propose the first multi-modality training strategy for joint training on multiple AQA, IQA, VQA, and AVQA databases. In order to overcome the issue of varying perceptual scales among different databases, our multi-modality training strategy first pretrains three feature extraction modules via the database-specific regression modules and then optimize the whole model using relative ranking information instead of MOSs. Finally, extensive experimental results show that our proposed model UNQA outperforms SOTA database-specific methods and is more parameter-efficient. 

\bibliographystyle{IEEEtran}
\bibliography{main}


\end{document}